\documentclass[english,twocolumn,aps,pra,showpacs, nofootinbib]{revtex4-1}

\usepackage[T1]{fontenc}
\usepackage[latin9]{inputenc}
\usepackage{amsmath,latexsym}
\usepackage{graphicx}
\usepackage[normalem]{ulem}
\usepackage{esint}
\usepackage{color}
\usepackage{lipsum}
\usepackage{float}
\usepackage[symbol]{footmisc}
\usepackage{amssymb}

\usepackage[colorlinks = true,
            linkcolor = blue,
            urlcolor  = blue,
            citecolor = blue,
            anchorcolor = blue]{hyperref}

\begin{document}

\title{Controlling Non-Markovian Dynamics Using a Light-based Structured Environment}

\author{Daniel F. Urrego$^{1,\dagger}$, Jefferson Fl\'orez$^{1,\ddagger}$
Ji\v{r}\'{i} Svozil\'{i}k$^{1}$, Mayerlin Nu\~nez$^{1}$, and  Alejandra Valencia$^{1,*}$}


\affiliation{$^{1}$ Laboratorio de Óptica Cuántica, Universidad de los Andes, A.A.
    4976, Bogotá D.C., Colombia}

\begin{abstract}
   We present the experimental control of Non-Markovian dynamics of open quantum systems simulated with photonic entities. The polarization of light is used as the system, whereas the surrounding environment is represented by light's spatial structure. The control of the dynamics is achieved by the engineering of the environment via spatial interference. Using the behavior of the trace distance, we are able to identify each dynamics and characterize the position of maximum revival of information.
\end{abstract}



\maketitle


\footnotetext[1]{ac.valencia@uniandes.edu.co}
\footnotetext[2]{df.urrego1720@uniandes.edu.co}
\footnotetext[3]{\textit{Currently at:} Department of Physics and Centre for Research in Photonics, University of Ottawa, 25 Templeton Street, Ottawa, Ontario K1N 6N5, Canada.}

\section{Introduction}

In Nature, real physical systems are affected by  their environments. In general, such physical systems can interchange with its environment particles, energy or information. One of the principal effects of an environment  is the loss of coherence in the system, or in other words, a loss of information codified in the system in favor of its surrounding~\cite{Schlosshauer2007}. When this is the case, the system undergoes what is called a Markovian dynamics. However, the flow of information does not need to be only one-directional: the information can be retrieved back to the system leading to what is called Non-Markovian dynamics~\cite{Breuer2007}. 

Markovian models are capable to describe dynamics of  many stochastic systems, such as, Brownian motion~\cite{Hu1992, Schlosshauer2008}, the current fluctuation in electric circuits~\cite{Rakos2008}, and chemical reactions~\cite{van1992}. On the other hand, Non-Markovian dynamics describes for example: light harvesting complexes coupled to their surroundings~\cite{Panitchayangkoon2010}, the emission of light from atoms or quantum dots coupled strongly to photonic crystals~\cite{Liu2017}, and a mechanical oscillator coupled to light~\cite{Groblacher2015}. 

Different experimental approaches have been devoted to show a controlled behavior of open quantum systems dynamics~\cite{Cimmarusti2015}. For example by using photonic platforms, in which discrete degrees of freedom usually play the role of the quantum system and the continuos variables constitute the environment, it has been possible to generate Non-Markovian dynamics~\cite{Cialdi2011, Liu2013} and to observe the transition from Markovian to Non-Markovian~\cite{Liu2011}. 

In this paper, we report the experimental control of Non-Markovian dynamics.  In our implementation, we associate light's polarization to the quantum system and light's transverse momentum distribution to the environment. By coupling these two degrees of freedom, we simulate a dynamics in which the transverse spatial displacement of a beam plays the role of the temporal variable. Differently from others experimental implementation that use photonic platforms, we obtain Non-Markovian dynamics by structuring the environment via spatial interference of light~\cite{Florez2016}. The recognition of Non-Markovian dynamics is done by observing the non-monotonic behavior of the trace distance~\cite{Breuer2016, Nielsen2010} and by the measure $\mathcal{N}_D$, based on the positive slope of the trace distance. From our results, we characterize the time  in which the maximal retrieval of information occurs. This last capability is useful, in applications such as quantum teleportation~\cite{Laine2014} and quantum key distribution (QKD)~\cite{Vasile2011}, since it allows to know the precise moment in which the information can be retrieved with a high fidelity. In the case of QKD, it also allows to carry out the protocol with high security.


\section{Theoretical background} 

Consider the case in which the system is the polarization of light and the environment is its transverse momentum. The initial state of the system can be written as $|\Psi^{s}\rangle=\alpha|V\rangle+\beta|H\rangle$, where $|V\rangle$ ($|H\rangle$) represents the vertical (horizontal) polarization, and $\alpha,\beta\in \mathbb{C}$ and fulfill $|\alpha|^2+|\beta|^2=1$. The environment is represented by the transverse momentum of light, $\vec{q}=\{q_x,q_y\}$, and it can be express by $|\Psi^{e}\rangle=\int d\vec{q}f\left(\vec{q}\right)|\vec{q}\rangle$ where $f(\vec{q})$ is the transverse momentum distribution that is normalized,  $\int d\vec{q}|f\left(\vec{q}\right)|=1$.

In order to simulate the system's dynamics, the system and environment must be coupled. Due to the coupling, the global initial state of system and environment, $|\Psi^{se}\rangle=|\Psi^{s}\rangle\otimes|\Psi^{e}\rangle$, is transformed by a unitary operation $\hat{U}$. In particular, consider the case in which $\hat{U}$ is associated with the spatial displacement, $d_c$, of a polarized beam in the $y$-direction fulfilling 
\begin{equation}
\begin{aligned}
\hat{U}\left(d_c\right)|V, q_y\rangle & =  
e^{id_cq_y}|V, q_y\rangle,\\
\hat{U}\left(d_c\right)|H, q_y\rangle & = e^{-i(d_cq_y+\varphi)}|H, q_y\rangle,
\end{aligned}
\label{eq: Unitary}
\end{equation}
where $\varphi$ is a generic phase. The spatial displacement, $d_c$, can be considered as the parameter that mediates the temporal evolution. The transformation $\hat{U}(d_c)$ evolves the global initial state to
\begin{equation} \label{eq: Coupling}
\begin{aligned}
&|\Psi^{se}\left(d_c\right)\rangle=\hat{U}\left(d_c\right)|\Psi^{se}\rangle\\
&=\int dq_y f\left(q_y\right)\bigg(\alpha e^{i(d_cq_y+\varphi)}|V\rangle+\beta e^{-id_cq_y}|H\rangle \bigg)|q_y\rangle ,
\end{aligned}
\end{equation}
revealing that system and environment are not separable anymore.  

  \begin{figure}[t]
  	\includegraphics[width=88mm]{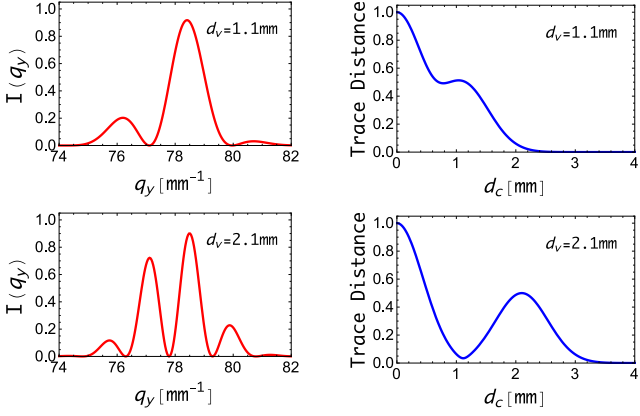}\\
  	\caption{Plots in the left column show different structured environments obtained by means of interferometric effect tuning the parameter $d_v$. Plots in the right column correspond to the evolution of the trace distance, for the environments on the left. Parameter $d_c$ that mediates the time.}
  	\label{fig: EnvironmentTracedistanceTheory}
  \end{figure}

In order to observe the effect of the environment over the system, it is necessary to trace out the environment variables. The density matrix of the final polarization state becomes,
\begin{equation}
\hat{\rho}^{s}\left(d_c\right)=\left(\begin{array}{cc}
|\alpha|^{2} & \alpha\beta^{*}\kappa(d_c)\\
\alpha^{*}\beta\kappa^*(d_c) & |\beta|^{2}
\end{array}\right),
\label{eq: DensityMatrix}
\end{equation}
where $\kappa(d_c)=\int dq_y |f\left(q_y\right)|^{2}e^{i(2q_y d_c+\varphi)}$. From Eq.~(\ref{eq: DensityMatrix}), it is clear  that the effect of the environment appears in the off-diagonal elements of the density matrix. This fact indicates that the environment induces decoherence.

The control of Non-Markovianity is achieved by structuring the environment. For that, the transverse momentum is engineered by means of spatial interference~\cite{Florez2016}. To this end, an input beam, with spatial distribution $f_{in}(q_y)$, is split in two parallel propagating beams separated by a distance $d_v$. By tuning $d_v$, it is posible to obtain a modulation in the environment given by $|f(q_y)|^2 \propto |f_{in}(q_y)|^2\left[1-\cos\left(2d_vq_y\right)\right]$. Considering $f_{in}(q_y)$ as a Gaussian beam, $f_{in}(q_y)  \propto e^\frac{-w_0^2( q_y-q_{0y})^2 }{4}$, the structured environment becomes 
\begin{equation}
|f\left(q_y \right)|^2  \propto
e^{-w_0^2\frac{\left(q_y-q_{0y}\right)^2}{2}}\left[1-\cos\left(2d_vq_y\right)\right],
\label{eq: Environment}
\end{equation}
where  $w_0$ is the waist of the input beam and $q_{y0}$ is the central transverse momentum. Examples of 2 different environments tuned by the parameter $d_v$ are shown in the left column of Fig.~\ref{fig: EnvironmentTracedistanceTheory}. The presence of modulation in these graphs clearly show the possibility of engineer different environments by means of spatial interference.
    
In order to identify the dynamics suffered by the system under the effect of an environment, two different initial states of the system, $\hat{\rho}^{s}_{1}$ and $\hat{\rho}^{s}_{2}$, are compared when they evolve during the same interval of time, $d_c$. This comparison can be done by means of the trace distance,
\begin{equation}
D\left(\hat{\rho}^{s}_{1}(d_c), \hat{\rho}^{s}_{2}(d_c) \right)=\frac{1}{2}Tr\left[\left| \hat{\rho}^{s}_{1}(d_c)-\hat{\rho}^{s}_{2}(d_c)\right| \right],
\label{eq: TraceDefinition}
\end{equation} 
that satisfies $0\leq D(d_c) \leq 1$ and has the maximum value when the two states, $\hat{\rho}^{s}_{1}$ and $\hat{\rho}^{s}_{2}$, are fully distinguishable. The behavior of $D(d_c)$ can be used to recognize if there is a Non-Markovian dynamics, since it corresponds to the case in which $D$ has a non-monotonical behavior as a function of time~ \cite{Breuer2009} i.e. as a function of $d_c$. 

It has been demonstrated  that two states  that maximized the trace distance have to be diametrically opposite in the Bloch sphere~\cite{Vasile2011} and need to have the highest values of the magnitude in the off-diagonal elements in the density matrix representation, Eq.~(\ref{eq: DensityMatrix}). Two possible states that satisfy such conditions are $\hat{\rho}^{s}_1=|\Psi_+^s\rangle\langle\Psi_+^s|$ and $\hat{\rho}^{s}_2=|\Psi_-^s\rangle\langle\Psi_-^s|$, where $|\Psi_{\pm}^s\rangle=\frac{1}{\sqrt{2}}(|V\rangle\pm|H\rangle)$, that lead to  
\begin{equation}
D\left(\hat{\rho}^{s}_{1}(d_c), \hat{\rho}^{s}_{2}(d_c) \right)= 
D\left(d_c\right)=|\kappa\left(d_c\right)|,
\label{eq: tracedistance}
\end{equation}
with
\begin{widetext}
\begin{equation}
|\kappa(d_c)|=\frac{e^{-\frac{2 d_c^2}{w_0^2}}}{1-e^{-\frac{2 d_v^2}{w_0^2}}\cos\left(2d_vq_{0y}\right)} \sqrt{e^{-\frac{4d_v^2}{w_0^2}}\left[\cos^2\left(2d_vq_{0y}\right)+\cosh^2\left(\frac{4 d_cd_v}{w_0^2}\right)-1\right]-2e^{-\frac{2d_v^2}{w_0^2}}\cos\left(2d_vq_{0y}\right)\cosh\left(\frac{4d_c d_v}{w_0^2}\right)+1}.
\label{eq: kappa}
\end{equation}
\end{widetext}

From Eq.~(\ref{eq: tracedistance}) and Eq.~(\ref{eq: kappa}), it is possible to see that the behavior of the trace distance depends on the characteristics of the environment defined by $d_v$.The right column of Fig.~\ref{fig: EnvironmentTracedistanceTheory} shows the behavior of the trace distance for the 2 different environments presented on the left side. The non-monotonic behavior of the trace distance is clearly seen, indicating that the dynamics induced by structured environments is Non-Markovian. 

\begin{figure}[H]
	
	\begin{tabular}{c}
		{\includegraphics[width=90mm]{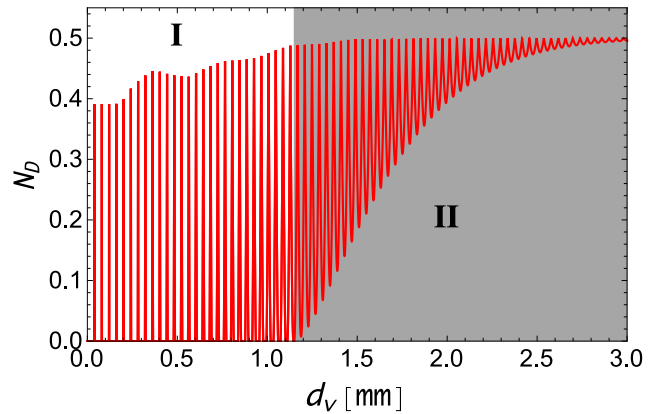}}\\
	\end{tabular}
	\caption{Theoretical measure $\mathcal{N}_D$ for different values of the parameter $d_v$ that defines the environments. Region $I$ corresponds to environments in which  $\mathcal{N}_D \ge 0$ indicating that there are dynamics that are Markovian and Non-Markovian. In region $II$, $\mathcal{N}_D > 0$ indicating that all the dynamics in this region are Non-Markovian.}
	\label{fig: DegreeMarkovianity}
\end{figure}

To quantify the Non-Markovianity of the dynamics,  we use the measure  $\mathcal{N}_D$, based on the trace distance and given by~\cite{Breuer2009, Breuer2016}:
\begin{equation}
\mathcal{N}_D=\mathrm{Max}\int_{\frac{d}{dy}>0}\frac{d}{dy}D\left(y\right)dy.
\label{eq: DegreeMarkovianity}
\end{equation}
This quantity takes a value bigger than 0 when a dynamics is Non-Markovian. Otherwise, $\mathcal{N}_D$ is not sufficient to make any inference about the type of dynamics~\cite{Breuer2016}. Mathematically, the fact that $\mathcal{N}_D$ reaches values equal to zero reflects that the first derivative of $|\kappa(d_c)|$ with respect to $d_c$ is not positive for some values of $d_v$. In Fig.~\ref{fig: DegreeMarkovianity}, we show $\mathcal{N}_D$ for different values of the parameter $d_v$ that defines the environments. This graph indicates that by structuring the environment, via spatial interference, it is possible to have situations in which the dynamics is Non-Markovian and others in which $\mathcal{N}_D=0$ and the dynamics is not defined according to the measure $\mathcal{N}_D$. However,  $\mathcal{N}_D=0$ is obtained for a Gaussian environment coupled to the system by means of a coupling that can be seen as a dephasing channel~\cite{Urrego2018}. For such channels, it has been demonstrated~\cite{Gong2015} that the evolution of the system satisfies a Lindbladian master equation with all the characteristics that correspond to a Markovian dynamics guaranteeing that, in our case, $\mathcal{N}_D=0$ is a sufficient condition to claim Markovianity. Since mathematically, the fact that $\mathcal{N}_D$ reaches values equal to zero reflects that the first derivative of $|\kappa(d_c)|$ with respect to $d_c$ is not positive for some values of $d_v$, in our case we can use the monotonic behavior of the trace distance to recognize a Markovian dynamics.

The presence of the oscillations in Fig.~\ref{fig: DegreeMarkovianity} is due to the dependence of $\cos(2d_vq_{0y})$ in $|k(d_c)|$, as can be seen in Eq.~(\ref{eq: kappa}). Two region, $I$ and $II$, can be identified in Fig.~\ref{fig: DegreeMarkovianity}. The frontier between both regions corresponds to $d_v \approx 1.15~mm$, the value at which the denominator  of Eq.~(\ref{eq: kappa}) becomes one and the hyperbolic functions dominate. In region $I$ and $II$, $\mathcal{N}_D$ can be equal or different than zero, indicating that some environments originate Markovian and and others Non-Markovian dynamics. In region $II$, the amplitude of the oscillations of $\mathcal{N}_D$ decreases and $\mathcal{N}_D$ tends to a constant value of 0.5 when $d_v$ increases. Differently to region $I$, region $II$ corresponds to environments for which $\mathcal{N}_D>0$, reveling that they induce an Non-Markovian dynamics.

\hspace{1 cm}


\onecolumngrid
\begin{center}
\begin{figure}[H]
\centering
\includegraphics[width=16cm]{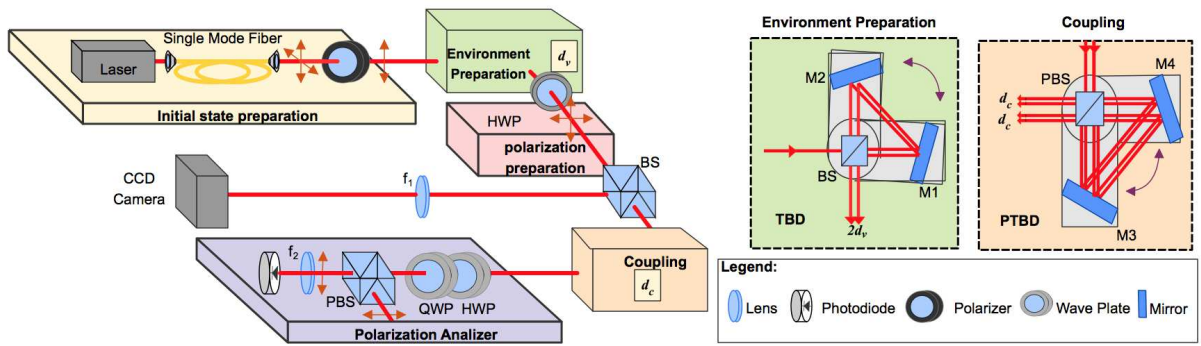}
\label{Fig:1}
\caption{Experimental setup composed of five steps. The first step is the preparation of a Gaussian spatial mode. The second step is the structuring of the environment, inset environment preparation. The third step is the polarization preparation. The forth step is the coupling of system and environment, inset coupling. Finally, the fifth step is the polarization analyzer.}
\label{fig: experimentalsetup}
\end{figure}
\end{center}
\twocolumngrid

\section{Experiment}

The experimental setup used to control the Non-Markovian dynamics is presented in Fig.~\ref{fig: experimentalsetup}. Five steps can be clearly recognized. The first step is the light source in which a $808$~nm CW laser (Thorlabs, CPS808) is coupled into a single mode fiber to obtain a Gaussian beam with a waist around  $w_0=0.88$~mm that can be considered collimated during the whole path of the experiment. A polarizer is used to set the light with vertical polarization. The second step, environment preparation, corresponds to the stage in which the environment is structured via spatial interference. As depicted in the inset environment preparations, the interference is produced by a tunable beam displacer (TBD) that separates an incoming beam into two parallel propagating beams separated by a distance $2d_v$. The TBD is composed of a beam splitter (BS) and two mirrors, M1 and M2, placed on an L-shaped platform that is mounted on a rotational stage. By rotating this platform, the separation $d_v$ can be tuned to generate different interference patterns, $|f\left(q_y\right)|^2$, as given in Eq.~(\ref{eq: Environment}). To observe the structure of the environment, we use a BS to guide the light to a $2f$ system that consists of a lens with a focal length $f_1=750$~mm and a CCD camera (ST-1603ME) placed in the Fourier plane. In the third step, a half wave plate (HWP) is used to define the initial state of the system, $\hat{\rho}^{s}$. Specifically, we prepared $\hat{\rho}^{s}_{1}$ and $\hat{\rho}^{s}_{2}$ by setting the HWP at $67.5^\circ$ and $22.5^\circ$, respectively. At this stage, system and environment are in a global initial state $|\Psi^{se}\rangle=|\Psi^{s}\rangle\otimes|\Psi^{e}\rangle$. Such prepared state enters to the fourth step of the experiment where the system is coupled to the environment by using a polarizing tunable beam displacer (PTBD). The PTBD has a similar structure to the TBD but uses a polarizing beam splitter (PBS) instead of a BS (inset coupling),  spliting an incoming beam into two parallel propagating beams with orthogonal polarization. For each beam that enters the PTBD, there are two output beams separated by the tunable distance $d_c$ that plays the role of time in our simulation of open quantum systems. The coupling performed by the PTBD corresponds to the operation $\hat{U}(d_c)$ in Eq.~(\ref{eq: Unitary}). Finally, in the fifth step, a polarization tomography analysis is implemented for different values of the separation $d_c$. This is done by sending the light through a HWP, a quarter wave plate (QWP) and a PBS. The light transmitted by the PBS is focused, with a lens $f_2$, into a photodiode (Thorlabs FDS100) while the light coming from the reflecting output of the PBS is neglected.


\begin{figure}[h]
\centering
\begin{tabular}{cc}
      \
      {\includegraphics[width=40mm]{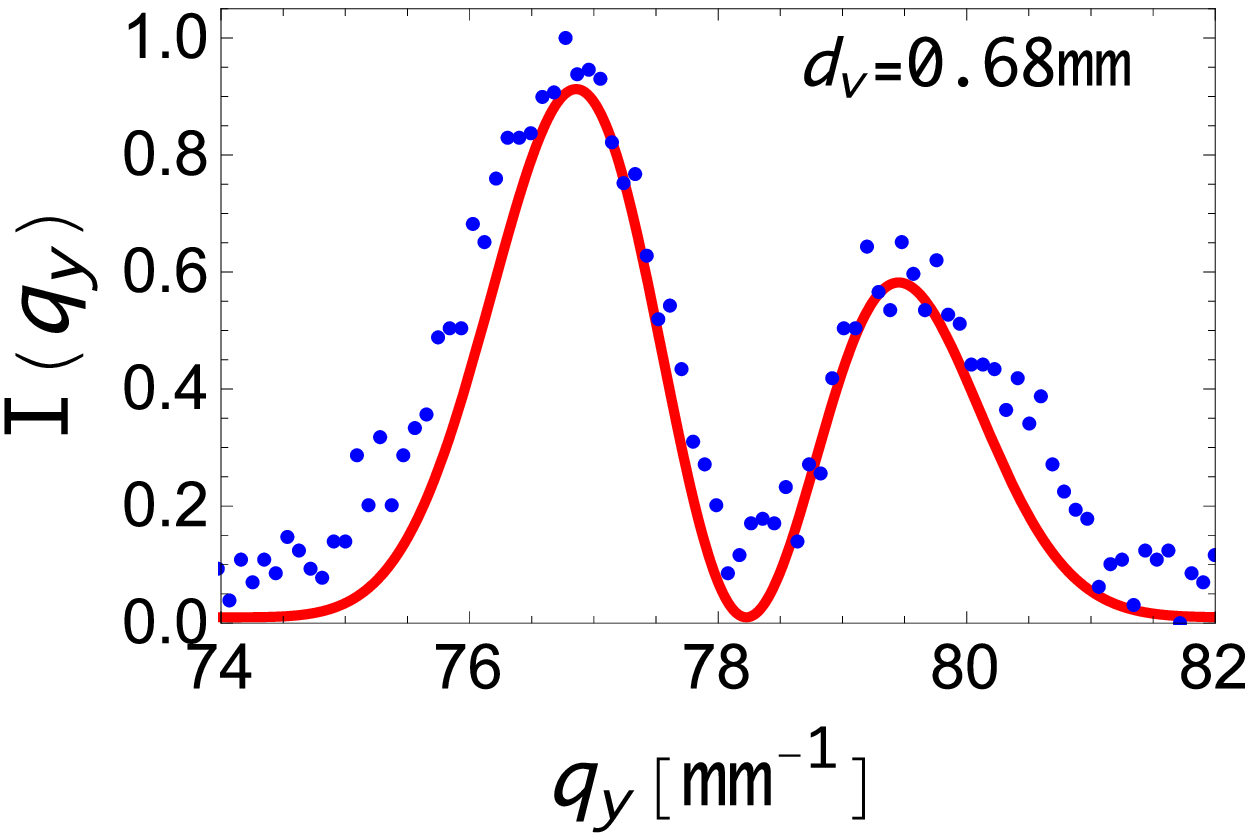}} &
      {\includegraphics[width=40mm]{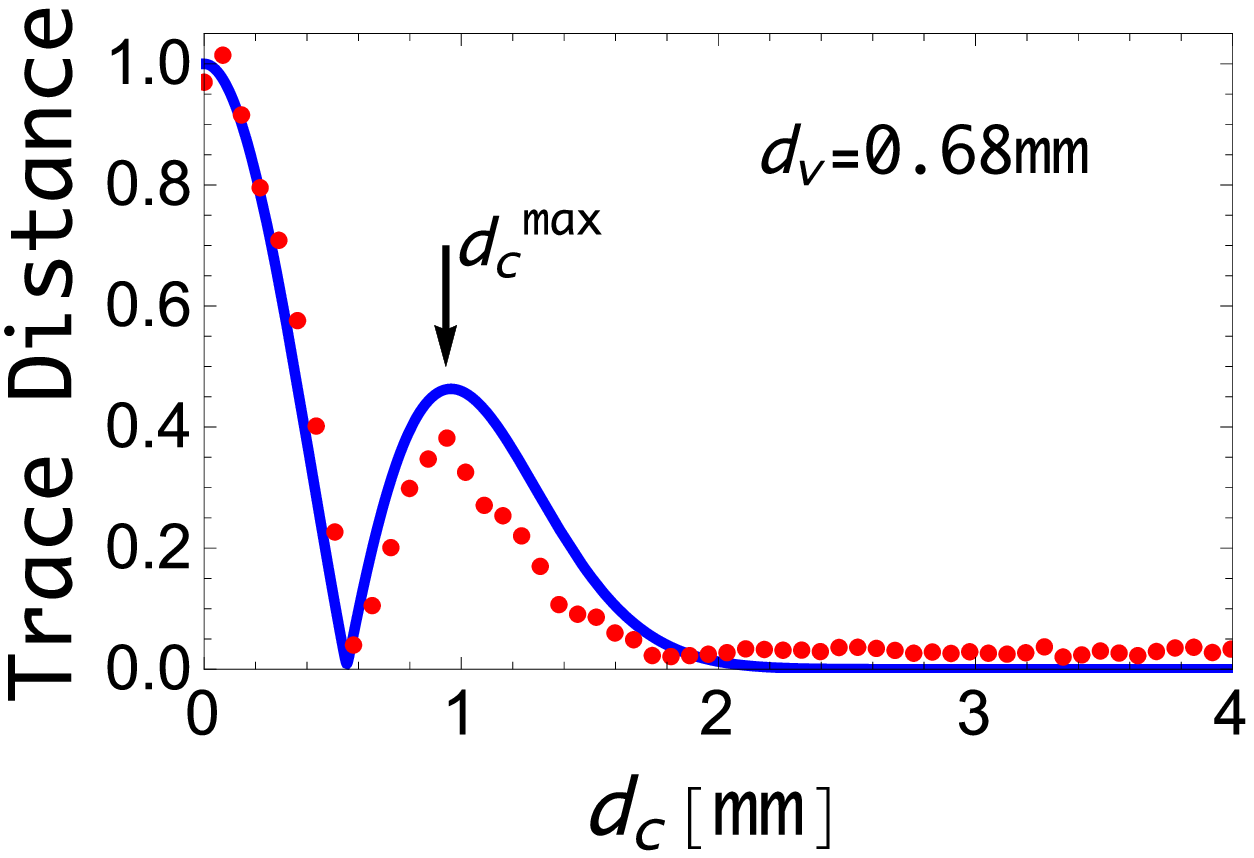}} \\
       \small a) &  b) \\
       {\includegraphics[width=40mm]{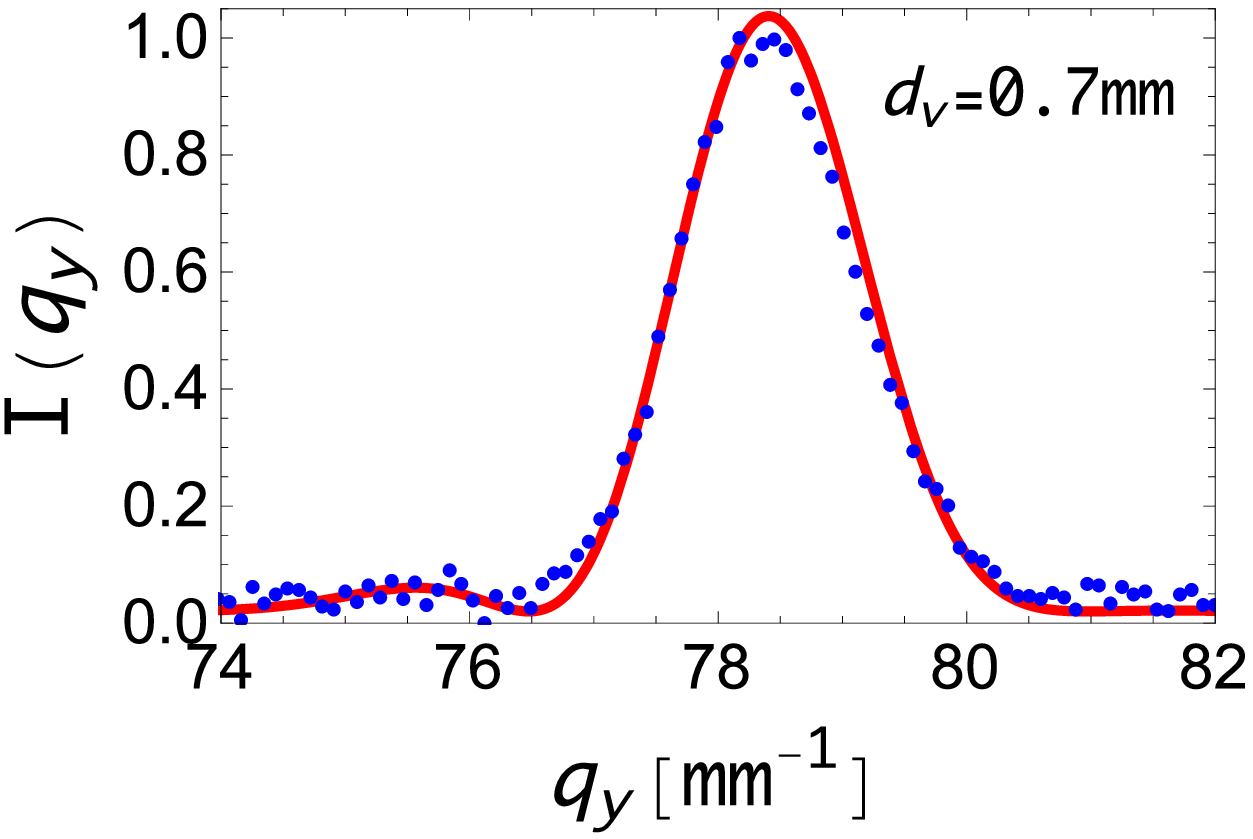}} &
      {\includegraphics[width=40mm]{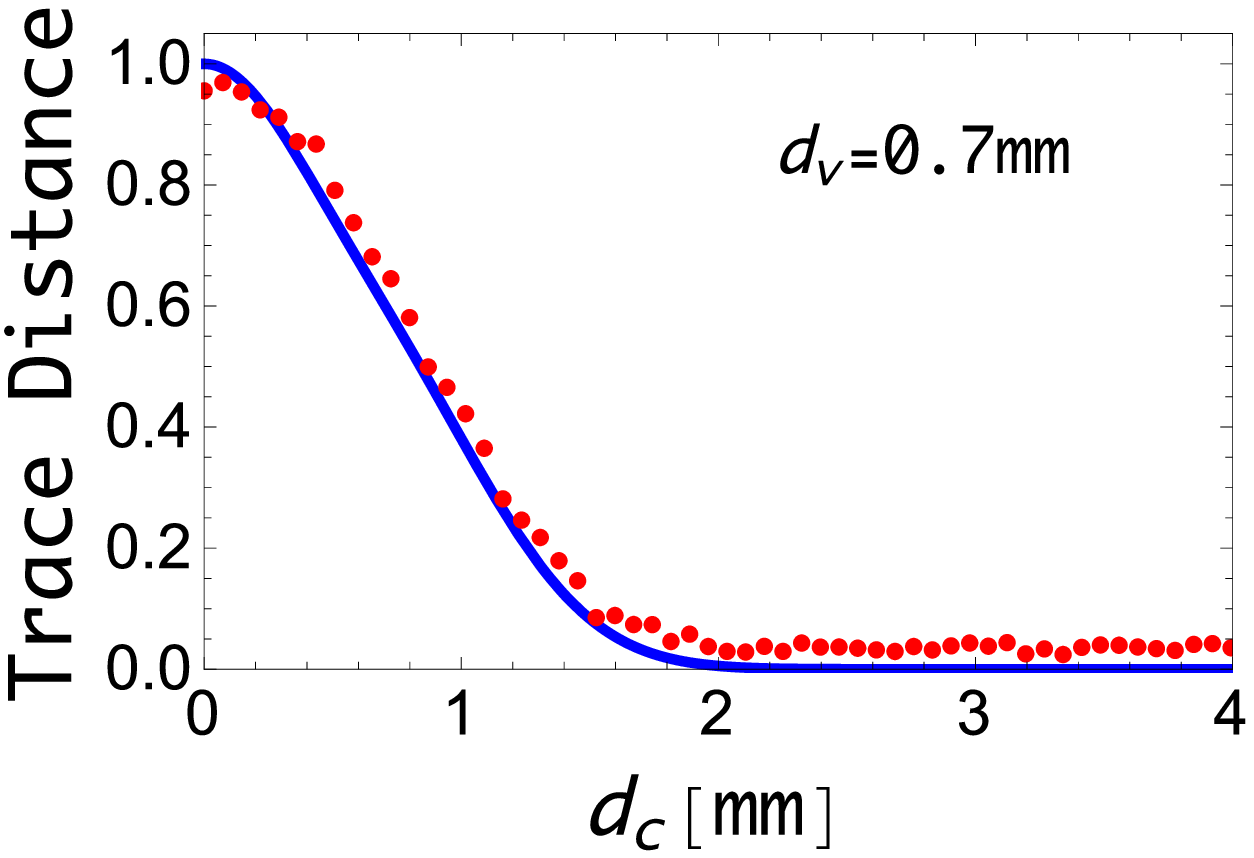}} \\
      \small c) &  d) \\
      {\includegraphics[width=40mm]{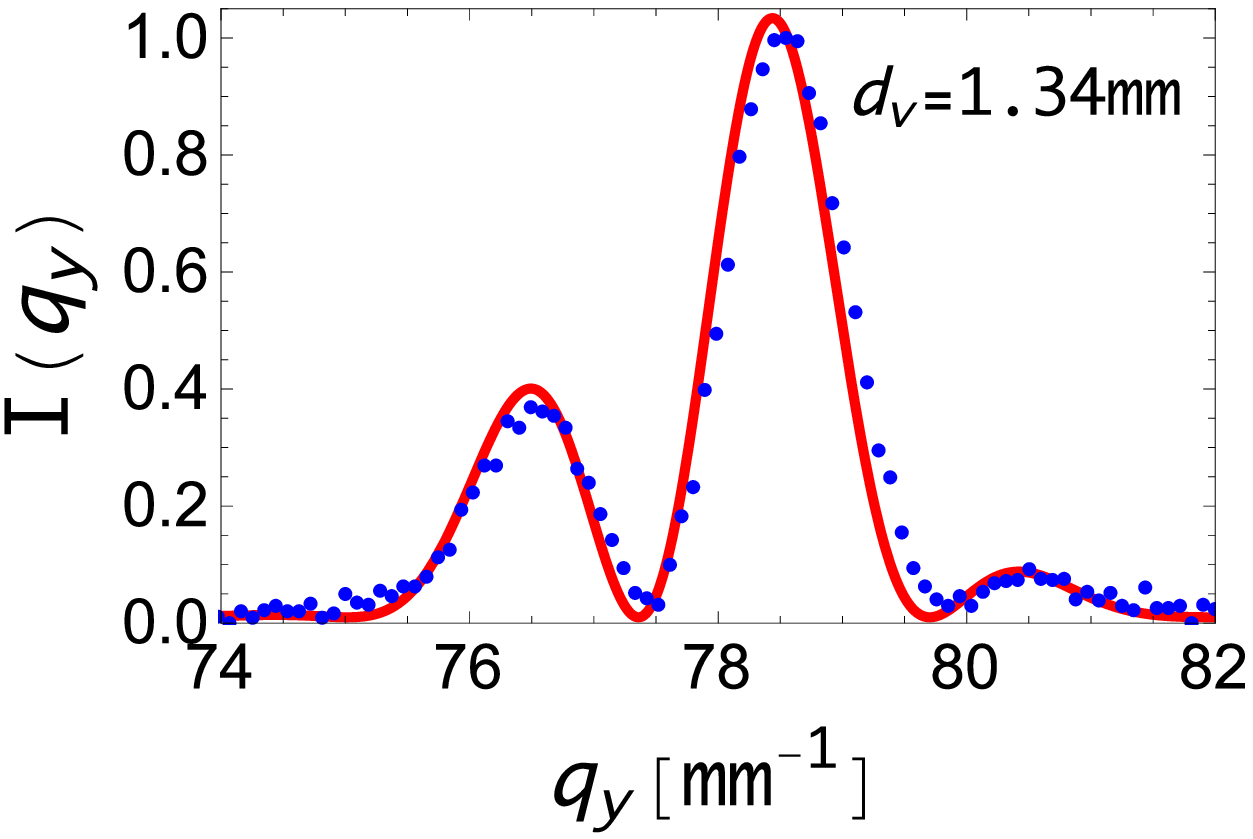}} &
      {\includegraphics[width=40mm]{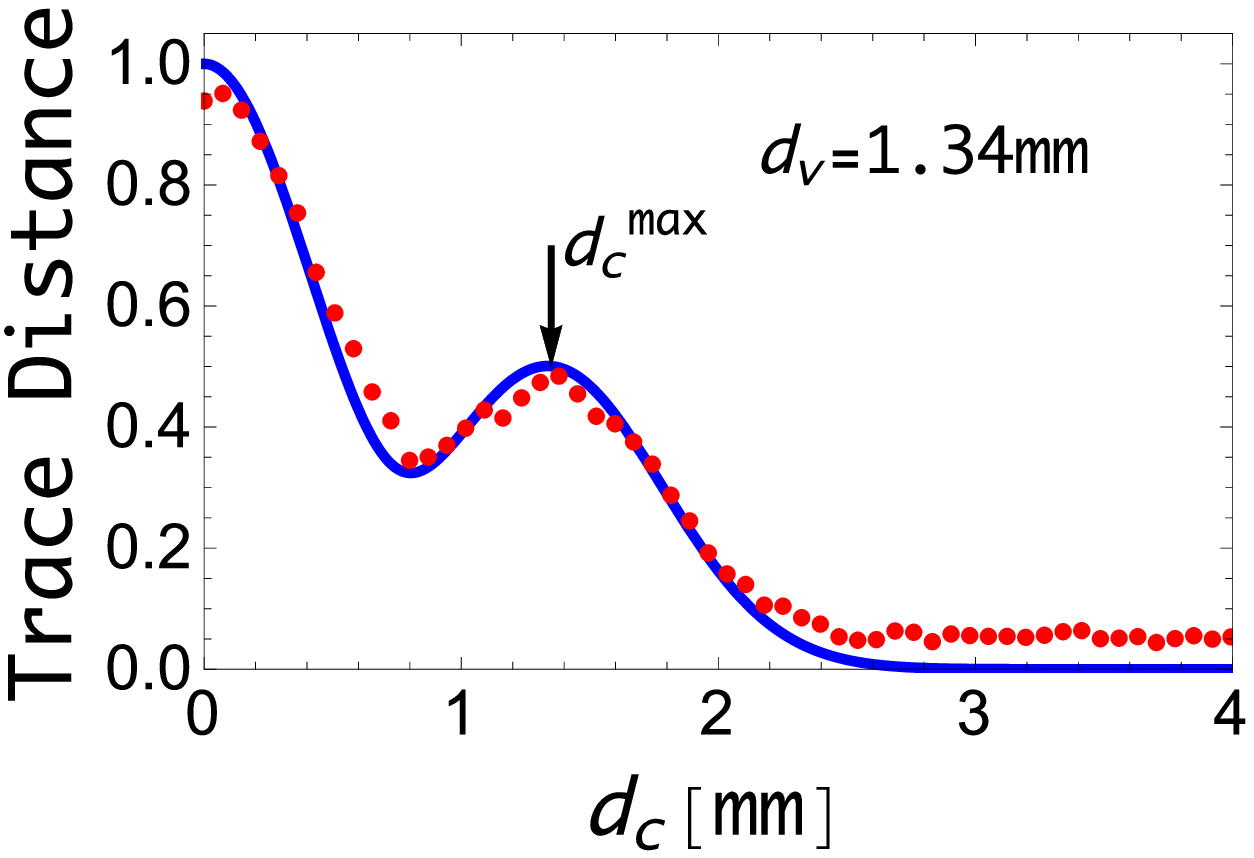}} \\
      \small e) &  f) \\
      {\includegraphics[width=40mm]{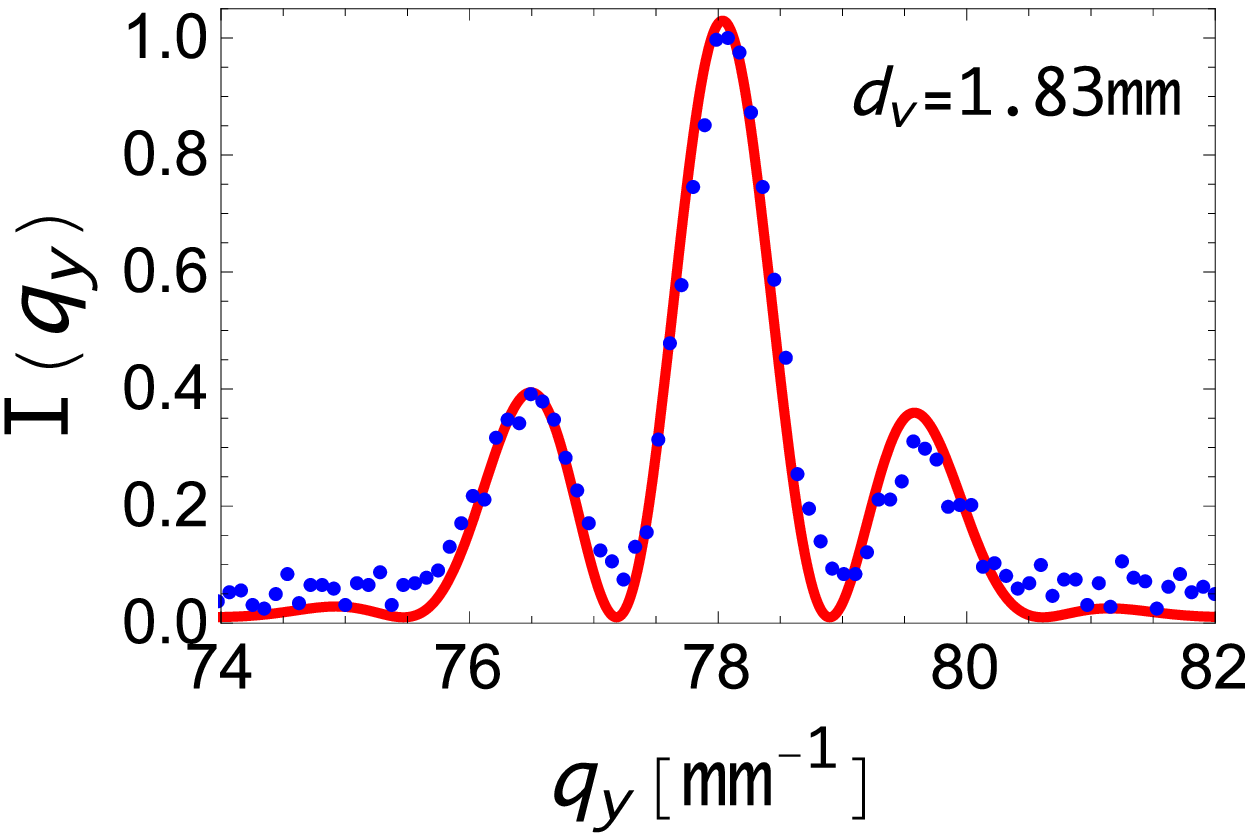}} &
      {\includegraphics[width=40mm]{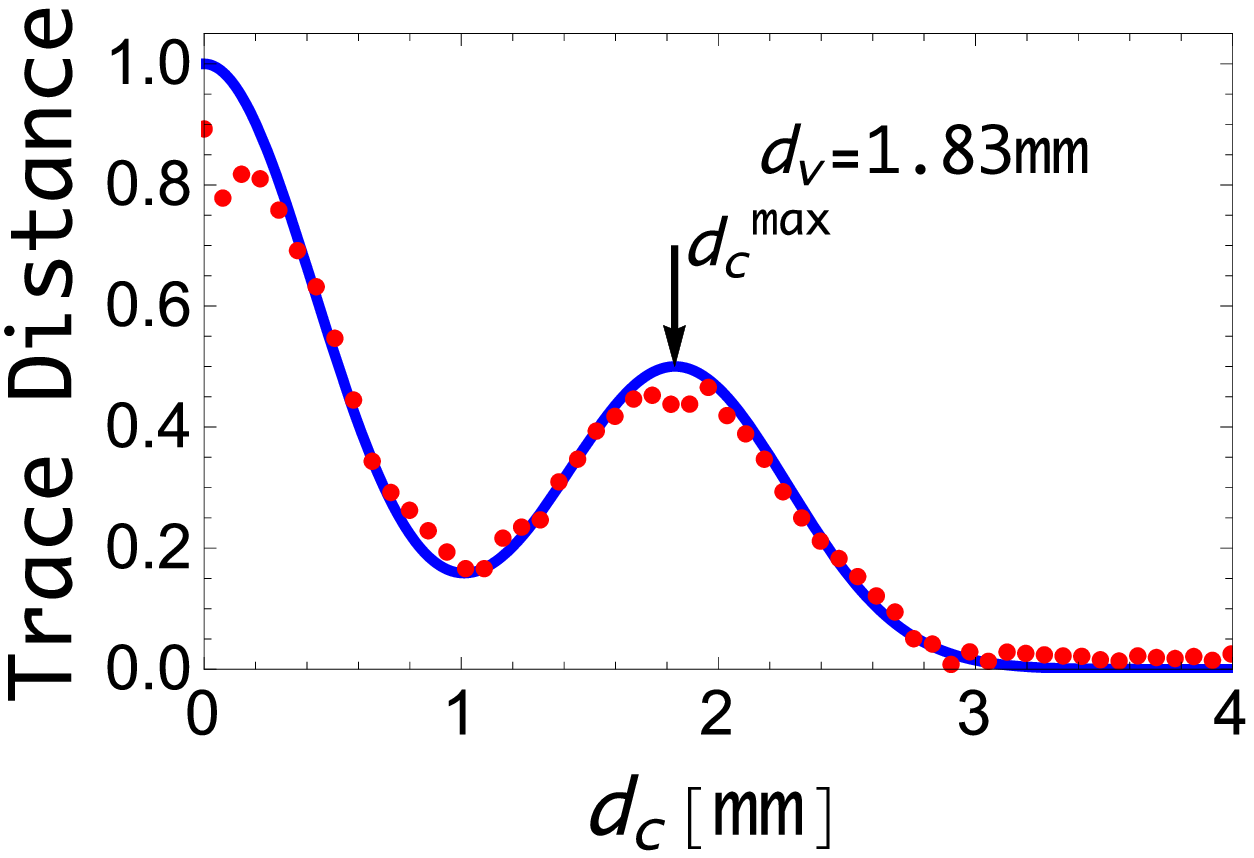}} \\
      \small g) &  h) \\
       {\includegraphics[width=40mm]{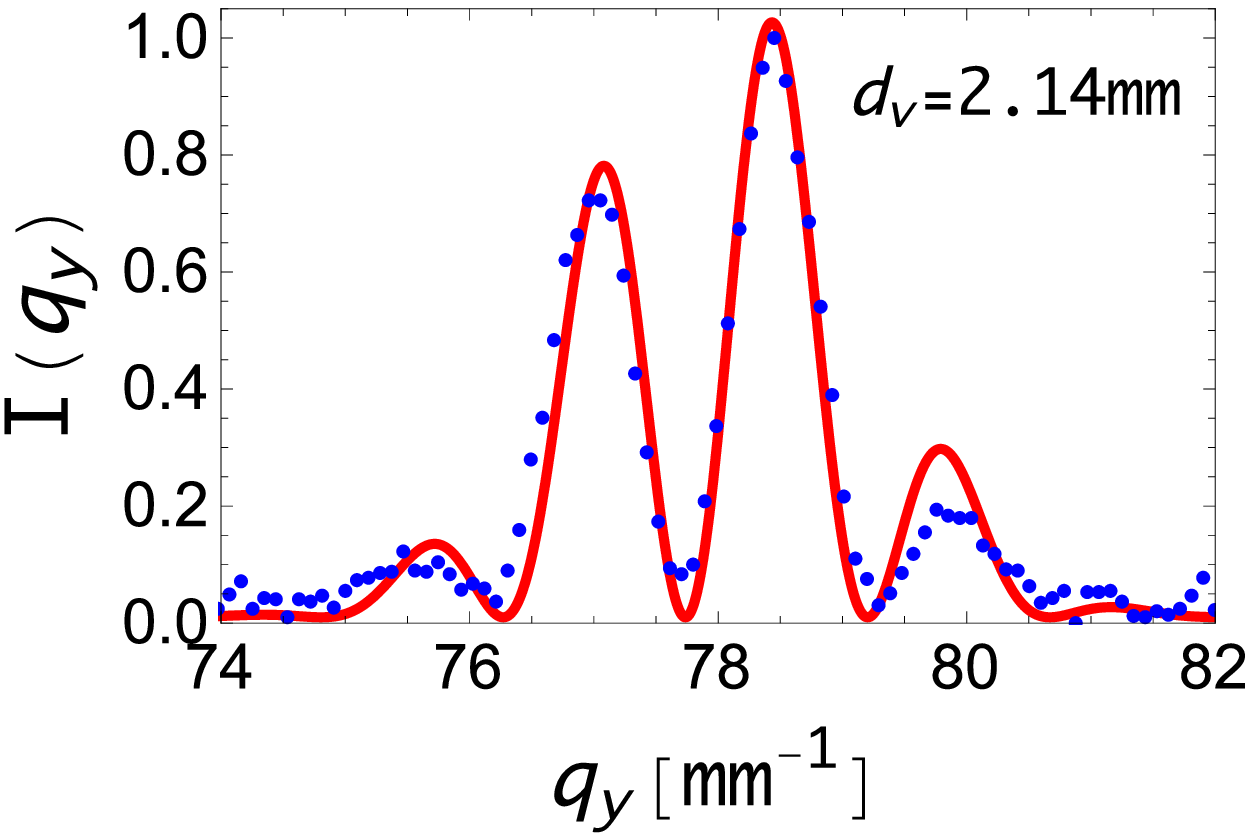}} &
      {\includegraphics[width=40mm]{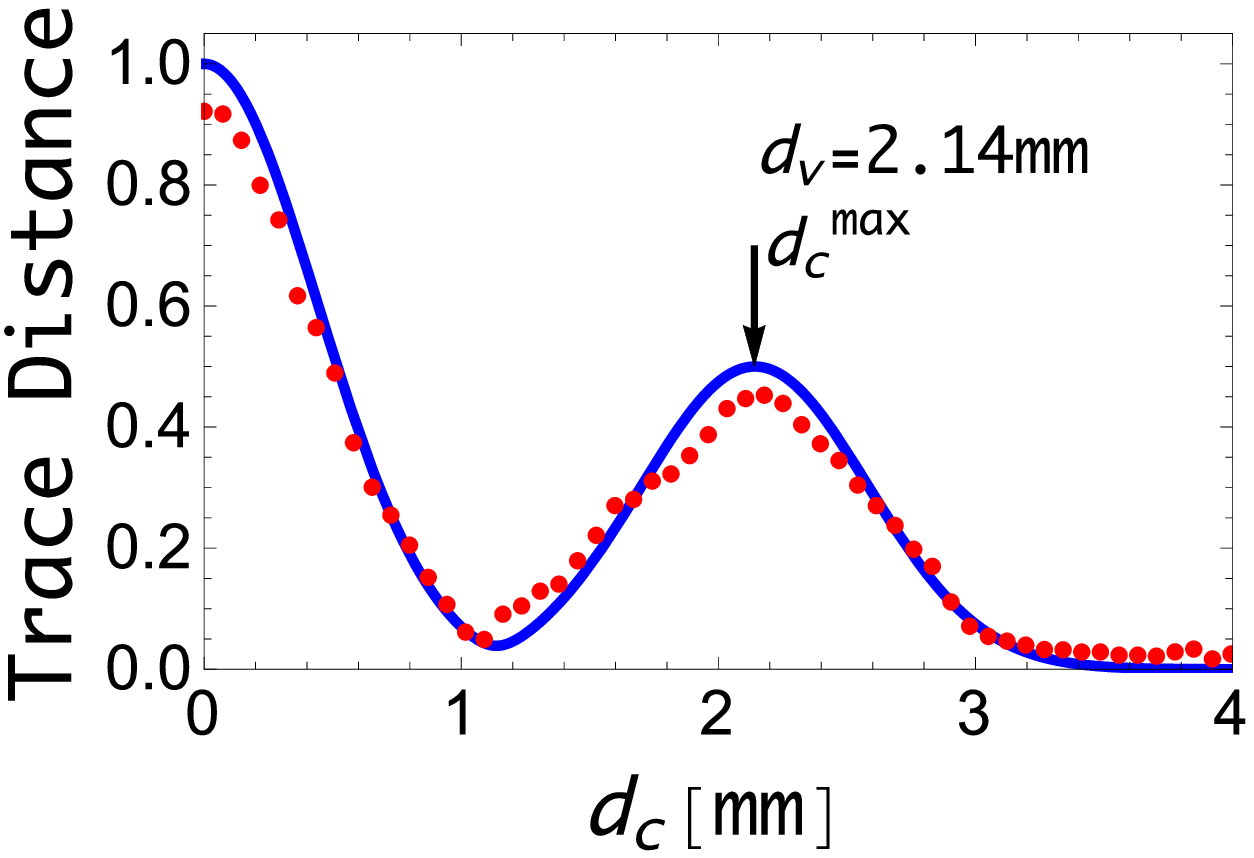}} \\   
      \small i) &  j) \\   
   
\end{tabular}
\caption{Plots in the left column show different structured environments obtained varying the parameter $d_v$. Plots in the right column correspond to the evolution of the trace distance for each environment on the left. Dots are experimental data and solid lines are the theoretical model.}
    \label{fig: experimentaldata}
\end{figure}

\section{Experimental Results}

We characterize the dynamics of an open quantum system for 5 different environments. The left column of Fig.~\ref{fig: experimentaldata} shows the environments obtained experimentally by tuning the parameter $d_v$ in the environment preparation stage. Solid lines corresponds to $|f(q_y)|^2$ as given by Eq.~(\ref{eq: Environment}) and dots are experimental data corresponding to a row of the data measured by the CCD camera, placed in the Fourier plane, with the background noise removed.

For each environment, we obtain the evolution of the trace distance by scanning over the parameter $d_c$ and performing the polarization tomography analysis for each of these values. In this way, we measure $\hat{\rho}_1^s(d_c)$ and $\hat{\rho}_2^s(d_c)$ and calcule the trace distance. The obtained evolution of the trace distance for each of the 5 different environment, reported above, is shown in the right column of Fig.~\ref{fig: experimentaldata}.  The dots are values obtained from experimental data and solid lines are their theoretical values calculated using  Eq.~(\ref{eq: tracedistance}) and Eq.~(\ref{eq: kappa}). From these plots, it is clear that when the environment presents a structure  (Fig.~\ref{fig: experimentaldata}(a, e, g, i)), i.e., it is composed by many peaks, the resulting evolution has a non-monotonic behavior indicating a Non-Markvovian dynamics. On the other hand, for the case in which the environment is nearly a Gaussian, Fig.~\ref{fig: experimentaldata}(c), the trace distance has a monotonic behavior  indicating  a Markovian dynamics.

The theoretical, $\mathcal{N}_D^{The}$, and experimental,  $\mathcal{N}_D^{Exp}$, values of  $\mathcal{N}_D$, defined in Eq.(\ref{eq: DegreeMarkovianity}), are shown in Table~\ref{tab: table1}. In all the cases the discrepancy between the theoretical and the experimental values is due to small fluctuations present in the experimental data mainly on the tails of the distributions. This discrepancy is particularly important in the case of the  Gaussian environment of Fig.~\ref{fig: experimentaldata}(c) because $\mathcal{N}_D^{Exp}$ leads to a wrong interpretation of the type of dynamics. However, for the rest of the environments, the value of $\mathcal{N}_D^{Exp}$ reveals the Non-Markovian feature of the dynamics.

\begin{table}[h]
\caption{\label{tab: table1}%
Measure $\mathcal{N}_D$
}
\begin{tabular}{ l  c c }
\hline
\hline
\textrm{Environment}& $\mathcal{N}_D^{The}$ & $\mathcal{N}_D^{Exp}$ \\
\hline
\hline
$d_v=0.68$~mm & 0.49& 0.40\\
$d_v=0.70$~mm & 0.00 & 0.09 \\
$d_v=1.34$~mm & 0.17 & 0.21\\
$d_v=1.84$~mm & 0.34 & 0.41\\
$d_v=2.14$~mm  & 0.46 & 0.42 \\
\hline
\hline
\end{tabular}
\end{table}
\hspace{1cm}
\begin{figure}[h]
\centering
\begin{tabular}{c}
      {\includegraphics[width=88mm]{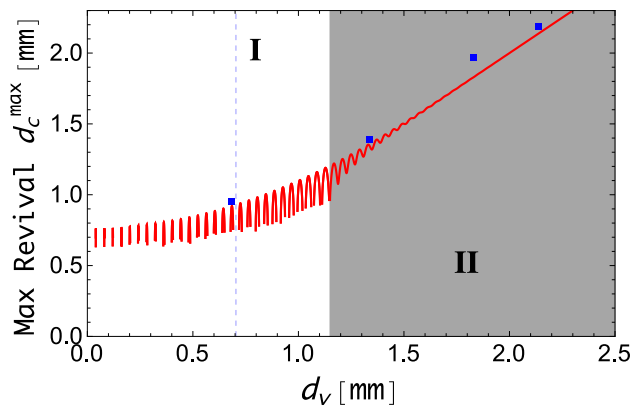}}\\
      \end{tabular}
\caption{Plots show, for different values of $d_v$ that defines the environment, the instant at which the maximum recovery of information occurs.}
    \label{fig: dcmaxExperimental}
\end{figure}

Another feature that can be recognized from the data of  the right column of Fig.~\ref{fig: experimentaldata} is the instant at which the maximum recovery of information, from the environment to the system, occurs. This instant for each environment is denoted by $d_{c}^{max}$ and it is indicated by an arrow. From the graphs, one observes that $d_{c}^{max}$ depends on the environment. The theoretical dependence of  $d_{c}^{max}$ with $d_v$, the parameter that characterizes the environment, is calculated by finding the local maximum of Eq.~(\ref{eq: kappa}) and it is depicted in  Fig.~\ref{fig: dcmaxExperimental}. Solid line is obtained from a computational model and squares are the experimental values that correspond to environment of Fig.~\ref{fig: experimentaldata}(a), Fig.~\ref{fig: experimentaldata}(e), Fig.~\ref{fig: experimentaldata}(g) and Fig.~\ref{fig: experimentaldata}(i). The dash line indicates the environment that is nearly a gaussian, Fig.~\ref{fig: experimentaldata}(c), for which the trace distance has a monotonic behavior; i.e. the dynamics is Markovian. The regions $I$ and $II$ are the same presented for the measure $\mathcal{N}_D$ of Non-Markovianity. In region $I$,  $d_c^{max}$ presents a discontinuous oscillatory behavior for the same environments in which $\mathcal{N}_D$ is equal to 0. In region $II$,  $d_{c}^{max}$ has a damping oscillation until it becomes a straight line indicating that $d_c^{max}$ is proportional to $d_v$.


\section{Conclusion}
 
 We reported the control of Non-Markovian dynamics by generating structured environment by means of spatial interferometric effects of light. In our experimental implementation the transverse separation between two light beams plays the role of time. We characterized the dynamics by observing a clear deviation from a monotonic behavior in the evolution of the trace distance and by obtaining the measure $\mathcal{N}_D$. Additionally, from our data, we were able to identify the moment in which the maximum recovery of information, previously lost in the environment, occurs. This fact is beneficial when considering environment-assisted effects that can be useful in applications.


\section{Acknowledgement}

Authors acknowledge financial support from Facultad de Ciencias, Universidad de los Andes. MNP acknowledges support from FAPA project, Universidad de los Andes. JS thanks the project 17-23005Y of the Czech Science Foundation. The Authors thank Prof. Luis Quiroga, Omar Calder\'{o}n-Losada and Juan Jose Mendoza for enlightening discussions.

\hspace{0cm}

%

\bibliographystyle{apsrev4-1}
\bibliography{Biblioteca.bib}

\begin{thebibliography}{21}%
\makeatletter
\providecommand \@ifxundefined [1]{%
 \@ifx{#1\undefined}
}%
\providecommand \@ifnum [1]{%
 \ifnum #1\expandafter \@firstoftwo
 \else \expandafter \@secondoftwo
 \fi
}%
\providecommand \@ifx [1]{%
 \ifx #1\expandafter \@firstoftwo
 \else \expandafter \@secondoftwo
 \fi
}%
\providecommand \natexlab [1]{#1}%
\providecommand \enquote  [1]{``#1''}%
\providecommand \bibnamefont  [1]{#1}%
\providecommand \bibfnamefont [1]{#1}%
\providecommand \citenamefont [1]{#1}%
\providecommand \href@noop [0]{\@secondoftwo}%
\providecommand \href [0]{\begingroup \@sanitize@url \@href}%
\providecommand \@href[1]{\@@startlink{#1}\@@href}%
\providecommand \@@href[1]{\endgroup#1\@@endlink}%
\providecommand \@sanitize@url [0]{\catcode `\\12\catcode `\$12\catcode
  `\&12\catcode `\#12\catcode `\^12\catcode `\_12\catcode `\%12\relax}%
\providecommand \@@startlink[1]{}%
\providecommand \@@endlink[0]{}%
\providecommand \url  [0]{\begingroup\@sanitize@url \@url }%
\providecommand \@url [1]{\endgroup\@href {#1}{\urlprefix }}%
\providecommand \urlprefix  [0]{URL }%
\providecommand \Eprint [0]{\href }%
\providecommand \doibase [0]{http://dx.doi.org/}%
\providecommand \selectlanguage [0]{\@gobble}%
\providecommand \bibinfo  [0]{\@secondoftwo}%
\providecommand \bibfield  [0]{\@secondoftwo}%
\providecommand \translation [1]{[#1]}%
\providecommand \BibitemOpen [0]{}%
\providecommand \bibitemStop [0]{}%
\providecommand \bibitemNoStop [0]{.\EOS\space}%
\providecommand \EOS [0]{\spacefactor3000\relax}%
\providecommand \BibitemShut  [1]{\csname bibitem#1\endcsname}%
\let\auto@bib@innerbib\@empty
\bibitem [{\citenamefont {Schlosshauer}(2007)}]{Schlosshauer2007}%
  \BibitemOpen
  \bibfield  {author} {\bibinfo {author} {\bibfnamefont {M.}~\bibnamefont
  {Schlosshauer}},\ }\href@noop {} {\emph {\bibinfo {title} {Decoherence: And
  the Quantum-To-Classical Transition}}},\ The Frontiers Collection\ (\bibinfo
  {publisher} {Springer},\ \bibinfo {year} {2007})\BibitemShut {NoStop}%
\bibitem [{\citenamefont {Breuer}\ and\ \citenamefont
  {Petruccione}(2007)}]{Breuer2007}%
  \BibitemOpen
  \bibfield  {author} {\bibinfo {author} {\bibfnamefont {H.}~\bibnamefont
  {Breuer}}\ and\ \bibinfo {author} {\bibfnamefont {F.}~\bibnamefont
  {Petruccione}},\ }\href@noop {} {\emph {\bibinfo {title} {The Theory of Open
  Quantum Systems}}}\ (\bibinfo  {publisher} {OUP Oxford},\ \bibinfo {year}
  {2007})\BibitemShut {NoStop}%
\bibitem [{\citenamefont {Hu}\ \emph {et~al.}(1992)\citenamefont {Hu},
  \citenamefont {Paz},\ and\ \citenamefont {Zhang}}]{Hu1992}%
  \BibitemOpen
  \bibfield  {author} {\bibinfo {author} {\bibfnamefont {B.~L.}\ \bibnamefont
  {Hu}}, \bibinfo {author} {\bibfnamefont {J.~P.}\ \bibnamefont {Paz}}, \ and\
  \bibinfo {author} {\bibfnamefont {Y.}~\bibnamefont {Zhang}},\ }\href
  {\doibase 10.1103/PhysRevD.45.2843} {\bibfield  {journal} {\bibinfo
  {journal} {Phys. Rev. D}\ }\textbf {\bibinfo {volume} {45}},\ \bibinfo
  {pages} {2843} (\bibinfo {year} {1992})}\BibitemShut {NoStop}%
\bibitem [{\citenamefont {Schlosshauer}\ \emph {et~al.}(2008)\citenamefont
  {Schlosshauer}, \citenamefont {Hines},\ and\ \citenamefont
  {Milburn}}]{Schlosshauer2008}%
  \BibitemOpen
  \bibfield  {author} {\bibinfo {author} {\bibfnamefont {M.}~\bibnamefont
  {Schlosshauer}}, \bibinfo {author} {\bibfnamefont {A.~P.}\ \bibnamefont
  {Hines}}, \ and\ \bibinfo {author} {\bibfnamefont {G.~J.}\ \bibnamefont
  {Milburn}},\ }\href {\doibase 10.1103/PhysRevA.77.022111} {\bibfield
  {journal} {\bibinfo  {journal} {Phys. Rev. A}\ }\textbf {\bibinfo {volume}
  {77}},\ \bibinfo {pages} {022111} (\bibinfo {year} {2008})}\BibitemShut
  {NoStop}%
\bibitem [{\citenamefont {R{\'a}kos}\ and\ \citenamefont
  {Harris}(2008)}]{Rakos2008}%
  \BibitemOpen
  \bibfield  {author} {\bibinfo {author} {\bibfnamefont {A.}~\bibnamefont
  {R{\'a}kos}}\ and\ \bibinfo {author} {\bibfnamefont {R.~J.}\ \bibnamefont
  {Harris}},\ }\href {http://stacks.iop.org/1742-5468/2008/i=05/a=P05005}
  {\bibfield  {journal} {\bibinfo  {journal} {J. Stat. Mech.}\ }\textbf
  {\bibinfo {volume} {2008}},\ \bibinfo {pages} {05005} (\bibinfo {year}
  {2008})}\BibitemShut {NoStop}%
\bibitem [{\citenamefont {Van~Kampen}(1992)}]{van1992}%
  \BibitemOpen
  \bibfield  {author} {\bibinfo {author} {\bibfnamefont {N.}~\bibnamefont
  {Van~Kampen}},\ }\href@noop {} {\emph {\bibinfo {title} {Stochastic Processes
  in Physics and Chemistry}}},\ North-Holland Personal Library\ (\bibinfo
  {publisher} {Elsevier Science},\ \bibinfo {year} {1992})\BibitemShut
  {NoStop}%
\bibitem [{\citenamefont {Panitchayangkoon}\ \emph {et~al.}(2010)\citenamefont
  {Panitchayangkoon}, \citenamefont {Hayes}, \citenamefont {Fransted},
  \citenamefont {Caram}, \citenamefont {Harel}, \citenamefont {Wen},
  \citenamefont {Blankenship},\ and\ \citenamefont
  {Engel}}]{Panitchayangkoon2010}%
  \BibitemOpen
  \bibfield  {author} {\bibinfo {author} {\bibfnamefont {G.}~\bibnamefont
  {Panitchayangkoon}}, \bibinfo {author} {\bibfnamefont {D.}~\bibnamefont
  {Hayes}}, \bibinfo {author} {\bibfnamefont {K.~A.}\ \bibnamefont {Fransted}},
  \bibinfo {author} {\bibfnamefont {J.~R.}\ \bibnamefont {Caram}}, \bibinfo
  {author} {\bibfnamefont {E.}~\bibnamefont {Harel}}, \bibinfo {author}
  {\bibfnamefont {J.}~\bibnamefont {Wen}}, \bibinfo {author} {\bibfnamefont
  {R.~E.}\ \bibnamefont {Blankenship}}, \ and\ \bibinfo {author} {\bibfnamefont
  {G.~S.}\ \bibnamefont {Engel}},\ }\href {\doibase 10.1073/pnas.1005484107}
  {\bibfield  {journal} {\bibinfo  {journal} {Proc. Natl. Acad. Sci.}\ }\textbf
  {\bibinfo {volume} {107}},\ \bibinfo {pages} {12766} (\bibinfo {year}
  {2010})}\BibitemShut {NoStop}%
\bibitem [{\citenamefont {Liu}\ and\ \citenamefont {Houck}(2016)}]{Liu2017}%
  \BibitemOpen
  \bibfield  {author} {\bibinfo {author} {\bibfnamefont {Y.}~\bibnamefont
  {Liu}}\ and\ \bibinfo {author} {\bibfnamefont {A.~A.}\ \bibnamefont
  {Houck}},\ }\href {http://dx.doi.org/10.1038/nphys3834} {\bibfield  {journal}
  {\bibinfo  {journal} {Nat. Phys.}\ }\textbf {\bibinfo {volume} {13}},\
  \bibinfo {pages} {48} (\bibinfo {year} {2016})}\BibitemShut {NoStop}%
\bibitem [{\citenamefont {Gr{\"o}blacher}\ \emph {et~al.}(2015)\citenamefont
  {Gr{\"o}blacher}, \citenamefont {Trubarov}, \citenamefont {Prigge},
  \citenamefont {Cole}, \citenamefont {Aspelmeyer},\ and\ \citenamefont
  {Eisert}}]{Groblacher2015}%
  \BibitemOpen
  \bibfield  {author} {\bibinfo {author} {\bibfnamefont {S.}~\bibnamefont
  {Gr{\"o}blacher}}, \bibinfo {author} {\bibfnamefont {A.}~\bibnamefont
  {Trubarov}}, \bibinfo {author} {\bibfnamefont {N.}~\bibnamefont {Prigge}},
  \bibinfo {author} {\bibfnamefont {G.~D.}\ \bibnamefont {Cole}}, \bibinfo
  {author} {\bibfnamefont {M.}~\bibnamefont {Aspelmeyer}}, \ and\ \bibinfo
  {author} {\bibfnamefont {J.}~\bibnamefont {Eisert}},\ }\href
  {http://dx.doi.org/10.1038/ncomms8606} {\bibfield  {journal} {\bibinfo
  {journal} {Nat. Commun.}\ }\textbf {\bibinfo {volume} {6}},\ \bibinfo {pages}
  {7606} (\bibinfo {year} {2015})}\BibitemShut {NoStop}%
\bibitem [{\citenamefont {Cimmarusti}\ \emph {et~al.}(2015)\citenamefont
  {Cimmarusti}, \citenamefont {Yan}, \citenamefont {Patterson}, \citenamefont
  {Corcos}, \citenamefont {Orozco},\ and\ \citenamefont
  {Deffner}}]{Cimmarusti2015}%
  \BibitemOpen
  \bibfield  {author} {\bibinfo {author} {\bibfnamefont {A.~D.}\ \bibnamefont
  {Cimmarusti}}, \bibinfo {author} {\bibfnamefont {Z.}~\bibnamefont {Yan}},
  \bibinfo {author} {\bibfnamefont {B.~D.}\ \bibnamefont {Patterson}}, \bibinfo
  {author} {\bibfnamefont {L.~P.}\ \bibnamefont {Corcos}}, \bibinfo {author}
  {\bibfnamefont {L.~A.}\ \bibnamefont {Orozco}}, \ and\ \bibinfo {author}
  {\bibfnamefont {S.}~\bibnamefont {Deffner}},\ }\href {\doibase
  10.1103/PhysRevLett.114.233602} {\bibfield  {journal} {\bibinfo  {journal}
  {Phys. Rev. Lett.}\ }\textbf {\bibinfo {volume} {114}},\ \bibinfo {pages}
  {233602} (\bibinfo {year} {2015})}\BibitemShut {NoStop}%
\bibitem [{\citenamefont {Cialdi}\ \emph {et~al.}(2011)\citenamefont {Cialdi},
  \citenamefont {Brivio}, \citenamefont {Tesio},\ and\ \citenamefont
  {Paris}}]{Cialdi2011}%
  \BibitemOpen
  \bibfield  {author} {\bibinfo {author} {\bibfnamefont {S.}~\bibnamefont
  {Cialdi}}, \bibinfo {author} {\bibfnamefont {D.}~\bibnamefont {Brivio}},
  \bibinfo {author} {\bibfnamefont {E.}~\bibnamefont {Tesio}}, \ and\ \bibinfo
  {author} {\bibfnamefont {M.~G.~A.}\ \bibnamefont {Paris}},\ }\href {\doibase
  10.1103/PhysRevA.83.042308} {\bibfield  {journal} {\bibinfo  {journal} {Phys.
  Rev. A}\ }\textbf {\bibinfo {volume} {83}},\ \bibinfo {pages} {042308}
  (\bibinfo {year} {2011})}\BibitemShut {NoStop}%
\bibitem [{\citenamefont {Liu}\ \emph {et~al.}(2013)\citenamefont {Liu},
  \citenamefont {Cao}, \citenamefont {Huang}, \citenamefont {Li}, \citenamefont
  {Guo}, \citenamefont {Laine}, \citenamefont {Breuer},\ and\ \citenamefont
  {Piilo}}]{Liu2013}%
  \BibitemOpen
  \bibfield  {author} {\bibinfo {author} {\bibfnamefont {B.-H.}\ \bibnamefont
  {Liu}}, \bibinfo {author} {\bibfnamefont {D.-Y.}\ \bibnamefont {Cao}},
  \bibinfo {author} {\bibfnamefont {Y.-F.}\ \bibnamefont {Huang}}, \bibinfo
  {author} {\bibfnamefont {C.-F.}\ \bibnamefont {Li}}, \bibinfo {author}
  {\bibfnamefont {G.-C.}\ \bibnamefont {Guo}}, \bibinfo {author} {\bibfnamefont
  {E.-M.}\ \bibnamefont {Laine}}, \bibinfo {author} {\bibfnamefont {H.-P.}\
  \bibnamefont {Breuer}}, \ and\ \bibinfo {author} {\bibfnamefont
  {J.}~\bibnamefont {Piilo}},\ }\href {\doibase 10.1038/srep01781} {\bibfield
  {journal} {\bibinfo  {journal} {Sci. Rep.}\ }\textbf {\bibinfo {volume}
  {3}},\ \bibinfo {pages} {1781} (\bibinfo {year} {2013})}\BibitemShut
  {NoStop}%
\bibitem [{\citenamefont {Liu}\ \emph {et~al.}(2011)\citenamefont {Liu},
  \citenamefont {Li}, \citenamefont {Huang}, \citenamefont {Li}, \citenamefont
  {Guo}, \citenamefont {Laine}, \citenamefont {Breuer},\ and\ \citenamefont
  {Piilo}}]{Liu2011}%
  \BibitemOpen
  \bibfield  {author} {\bibinfo {author} {\bibfnamefont {B.-H.}\ \bibnamefont
  {Liu}}, \bibinfo {author} {\bibfnamefont {L.}~\bibnamefont {Li}}, \bibinfo
  {author} {\bibfnamefont {Y.-F.}\ \bibnamefont {Huang}}, \bibinfo {author}
  {\bibfnamefont {C.-F.}\ \bibnamefont {Li}}, \bibinfo {author} {\bibfnamefont
  {G.-C.}\ \bibnamefont {Guo}}, \bibinfo {author} {\bibfnamefont {E.-M.}\
  \bibnamefont {Laine}}, \bibinfo {author} {\bibfnamefont {H.-P.}\ \bibnamefont
  {Breuer}}, \ and\ \bibinfo {author} {\bibfnamefont {J.}~\bibnamefont
  {Piilo}},\ }\href {http://dx.doi.org/10.1038/nphys2085} {\bibfield  {journal}
  {\bibinfo  {journal} {Nat. Phys.}\ }\textbf {\bibinfo {volume} {7}},\
  \bibinfo {pages} {931} (\bibinfo {year} {2011})}\BibitemShut {NoStop}%
\bibitem [{\citenamefont {Fl{\'o}rez}\ \emph {et~al.}(2016)\citenamefont
  {Fl{\'o}rez}, \citenamefont {{\'A}lvarez}, \citenamefont
  {Calder{\'o}n-Losada}, \citenamefont {Salazar-Serrano},\ and\ \citenamefont
  {Valencia}}]{Florez2016}%
  \BibitemOpen
  \bibfield  {author} {\bibinfo {author} {\bibfnamefont {J.}~\bibnamefont
  {Fl{\'o}rez}}, \bibinfo {author} {\bibfnamefont {J.-R.}\ \bibnamefont
  {{\'A}lvarez}}, \bibinfo {author} {\bibfnamefont {O.}~\bibnamefont
  {Calder{\'o}n-Losada}}, \bibinfo {author} {\bibfnamefont {L.~J.}\
  \bibnamefont {Salazar-Serrano}}, \ and\ \bibinfo {author} {\bibfnamefont
  {A.}~\bibnamefont {Valencia}},\ }\href
  {http://stacks.iop.org/2040-8986/18/i=12/a=125201} {\bibfield  {journal}
  {\bibinfo  {journal} {J. Opt.}\ }\textbf {\bibinfo {volume} {18}},\ \bibinfo
  {pages} {125201} (\bibinfo {year} {2016})}\BibitemShut {NoStop}%
\bibitem [{\citenamefont {Breuer}\ \emph {et~al.}(2016)\citenamefont {Breuer},
  \citenamefont {Laine}, \citenamefont {Piilo},\ and\ \citenamefont
  {Vacchini}}]{Breuer2016}%
  \BibitemOpen
  \bibfield  {author} {\bibinfo {author} {\bibfnamefont {H.-P.}\ \bibnamefont
  {Breuer}}, \bibinfo {author} {\bibfnamefont {E.-M.}\ \bibnamefont {Laine}},
  \bibinfo {author} {\bibfnamefont {J.}~\bibnamefont {Piilo}}, \ and\ \bibinfo
  {author} {\bibfnamefont {B.}~\bibnamefont {Vacchini}},\ }\href {\doibase
  10.1103/RevModPhys.88.021002} {\bibfield  {journal} {\bibinfo  {journal}
  {Rev. Mod. Phys.}\ }\textbf {\bibinfo {volume} {88}},\ \bibinfo {pages}
  {021002} (\bibinfo {year} {2016})}\BibitemShut {NoStop}%
\bibitem [{\citenamefont {Nielsen}\ and\ \citenamefont
  {Chuang}(2011)}]{Nielsen2010}%
  \BibitemOpen
  \bibfield  {author} {\bibinfo {author} {\bibfnamefont {M.~A.}\ \bibnamefont
  {Nielsen}}\ and\ \bibinfo {author} {\bibfnamefont {I.~L.}\ \bibnamefont
  {Chuang}},\ }\href@noop {} {\emph {\bibinfo {title} {Quantum Computation and
  Quantum Information: 10th Anniversary Edition}}},\ \bibinfo {edition} {10th}\
  ed.\ (\bibinfo  {publisher} {Cambridge University},\ \bibinfo {year}
  {2011})\BibitemShut {NoStop}%
\bibitem [{\citenamefont {Laine}\ \emph {et~al.}(2014)\citenamefont {Laine},
  \citenamefont {Breuer},\ and\ \citenamefont {Piilo}}]{Laine2014}%
  \BibitemOpen
  \bibfield  {author} {\bibinfo {author} {\bibfnamefont {E.-M.}\ \bibnamefont
  {Laine}}, \bibinfo {author} {\bibfnamefont {H.-P.}\ \bibnamefont {Breuer}}, \
  and\ \bibinfo {author} {\bibfnamefont {J.}~\bibnamefont {Piilo}},\ }\href
  {http://dx.doi.org/10.1038/srep04620} {\bibfield  {journal} {\bibinfo
  {journal} {Sci. Rep.}\ }\textbf {\bibinfo {volume} {4}},\ \bibinfo {pages}
  {4620} (\bibinfo {year} {2014})}\BibitemShut {NoStop}%
\bibitem [{\citenamefont {Vasile}\ \emph {et~al.}(2011)\citenamefont {Vasile},
  \citenamefont {Olivares}, \citenamefont {Paris},\ and\ \citenamefont
  {Maniscalco}}]{Vasile2011}%
  \BibitemOpen
  \bibfield  {author} {\bibinfo {author} {\bibfnamefont {R.}~\bibnamefont
  {Vasile}}, \bibinfo {author} {\bibfnamefont {S.}~\bibnamefont {Olivares}},
  \bibinfo {author} {\bibfnamefont {M.~A.}\ \bibnamefont {Paris}}, \ and\
  \bibinfo {author} {\bibfnamefont {S.}~\bibnamefont {Maniscalco}},\ }\href
  {\doibase 10.1103/PhysRevA.83.042321} {\bibfield  {journal} {\bibinfo
  {journal} {Phys. Rev. A}\ }\textbf {\bibinfo {volume} {83}},\ \bibinfo
  {pages} {042321} (\bibinfo {year} {2011})}\BibitemShut {NoStop}%
\bibitem [{\citenamefont {Breuer}\ \emph {et~al.}(2009)\citenamefont {Breuer},
  \citenamefont {Laine},\ and\ \citenamefont {Piilo}}]{Breuer2009}%
  \BibitemOpen
  \bibfield  {author} {\bibinfo {author} {\bibfnamefont {H.-P.}\ \bibnamefont
  {Breuer}}, \bibinfo {author} {\bibfnamefont {E.-M.}\ \bibnamefont {Laine}}, \
  and\ \bibinfo {author} {\bibfnamefont {J.}~\bibnamefont {Piilo}},\ }\href
  {\doibase 10.1103/PhysRevLett.103.210401} {\bibfield  {journal} {\bibinfo
  {journal} {Phys. Rev. Lett.}\ }\textbf {\bibinfo {volume} {103}},\ \bibinfo
  {pages} {210401} (\bibinfo {year} {2009})}\BibitemShut {NoStop}%
\bibitem [{\citenamefont {Urrego}\ \emph {et~al.}(2018)\citenamefont {Urrego},
  \citenamefont {\'{A}lvarez}, \citenamefont {Calder\'{o}n-Losada},
  \citenamefont {Svozil\'{i}k}, \citenamefont {{n}ez},\ and\ \citenamefont
  {Valencia}}]{Urrego2018}%
  \BibitemOpen
  \bibfield  {author} {\bibinfo {author} {\bibfnamefont {D.~F.}\ \bibnamefont
  {Urrego}}, \bibinfo {author} {\bibfnamefont {J.-R.}\ \bibnamefont
  {\'{A}lvarez}}, \bibinfo {author} {\bibfnamefont {O.}~\bibnamefont
  {Calder\'{o}n-Losada}}, \bibinfo {author} {\bibfnamefont {J.}~\bibnamefont
  {Svozil\'{i}k}}, \bibinfo {author} {\bibfnamefont {M.~N.}\ \bibnamefont
  {{n}ez}}, \ and\ \bibinfo {author} {\bibfnamefont {A.}~\bibnamefont
  {Valencia}},\ }\href {\doibase 10.1364/OE.26.011940} {\bibfield  {journal}
  {\bibinfo  {journal} {Opt. Express}\ }\textbf {\bibinfo {volume} {26}},\
  \bibinfo {pages} {11940} (\bibinfo {year} {2018})}\BibitemShut {NoStop}%
\bibitem [{\citenamefont {ling Gong}\ \emph {et~al.}(2015)\citenamefont {ling
  Gong}, \citenamefont {Zhou},\ and\ \citenamefont {Schirmer}}]{Gong2015}%
  \BibitemOpen
  \bibfield  {author} {\bibinfo {author} {\bibfnamefont {E.}~\bibnamefont {ling
  Gong}}, \bibinfo {author} {\bibfnamefont {W.}~\bibnamefont {Zhou}}, \ and\
  \bibinfo {author} {\bibfnamefont {S.}~\bibnamefont {Schirmer}},\ }\href
  {\doibase https://doi.org/10.1016/j.physleta.2014.11.042} {\bibfield
  {journal} {\bibinfo  {journal} {Physics Letters A}\ }\textbf {\bibinfo
  {volume} {379}},\ \bibinfo {pages} {272 } (\bibinfo {year}
  {2015})}\BibitemShut {NoStop}%
\end{thebibliography}%

\end{document}